\documentclass[conference]{IEEEtran}
\IEEEoverridecommandlockouts
\usepackage{cite}
\usepackage{amsmath,amssymb,amsfonts}
\usepackage{algorithm}
\usepackage{algorithmic}
\usepackage{graphicx}
\usepackage{epstopdf}
\usepackage{mathrsfs}
\usepackage{verbatim}
\usepackage{textcomp}
\usepackage{xcolor}
\DeclareMathOperator{\sinc}{sinc} 
\DeclareMathOperator{\rect}{rect}

\begin{document}

\title{MIMO Beampattern Synthesis using Adaptive Frequency Modulated Waveforms\\
\thanks{This research was funded by the Naval Undersea Warfare Center's In-House Laboratory Independent Research (ILIR) program from the Office of Naval Research (ONR) under N0001424WX00177.}
}

\author{\IEEEauthorblockN{David A. Hague}
\IEEEauthorblockA{\textit{Naval Undersea Warfare Center} \\
1176 Howell St., Newport, RI, USA \\
david.a.hague.civ@us.navy.mil}
}

\maketitle

\begin{abstract}
This paper demonstrates a method that synthesizes narrowband Multiple-Input Multiple-Output (MIMO) beampatterns using the Multi-Tone Sinusoidal Frequency Modulated (MTSFM) waveform model.  MIMO arrays transmit unique waveforms on each of their elements which increases the degrees of freedom available to synthesize novel transmit beampatterns.  The MIMO beampattern shape is determined by the structure of the MIMO correlation matrix whose entries are the inner products between the waveforms transmitted on each element.  The MTSFM waveform possesses an instantaneous phase that is represented as a finite Fourier series.  The Fourier coefficients are modified to synthesize sets of waveforms whose correlation matrix realizes a desired MIMO transmit beampattern.  The MIMO correlation matrix for a MTSFM waveform set has an analytical form expressed in terms of Generalized Bessel Functions.  These mathematical properties are utilized to develop an optimization routine that synthesizes MTSFM waveform sets to approximate a desired MIMO transmit beampattern.  The performance of this optimization routine is then demonstrated via an illustrative design example.   
\end{abstract}


\section{Introduction}
\label{sec:intro}
Multiple-Input Multiple-Output (MIMO) arrays have seen extensive use in communications and radar applications \cite{JianLiBookI}.  A coherent MIMO array is a generalization of the standard phased-array which transmits a unique waveform on each element.  This increases the number of degrees of freedom in the transmit array design which facilitates synthesizing novel transmit beampattern shapes.  The inner products between each of the waveforms in the waveform set, also known as the correlation matrix, determines the structure of the resulting transmit beampattern of the MIMO array.  The beampattern synthesis problem is typically a two step process.  The first step is to find a correlation matrix that synthesizes a close fit to a desired beampattern shape.  The second step is then designing a waveform set that approximates that correlation matrix.  There exists a rich literature describing methods that find the correlation matrix that realizes a desired MIMO transmit beampattern, see \cite{MIMO_SanAntonio, MIMO_LI} for an overview of these methods.  There are also many efforts which focus on finding waveform sets that possess a specific correlation matrix structure, see \cite{MIMO_Overview, MIMO_Waveform} and references therein for an overview of these methods.  

While a complete overview of the MIMO waveform design literature is not possible here, there are several specific research efforts and practical design considerations that inspired this work.  Fuhrmann and San Antonio \cite{MIMO_SanAntonio} established algorithms for synthesizing both the desired correlation matrix and waveforms that closely approximate the correlation matrix using Binary Phase Shift-Keyed (BPSK) waveforms.  Li \emph{et. al.} \cite{MIMO_Li_CAN} developed a series of computationally efficient algorithms that produced constant modulus waveform sets to realize a desired MIMO beampattern.  There continues to be similar efforts focused on computationally efficient synthesis methods for MIMO waveform set design \cite{RangaswamyI, PalomarI, Prabhu_Babu_I, Prabhu_Babu_II}.  In addition to realizing a desired beampattern shape, the waveform sets should also possess a constant modulus and a compact spectral shape.  The constant modulus property ensures the waveform is amenable to efficient transmission on practical power amplifiers.  Spectrally compact waveforms allow for more efficient operation in spectrally crowded environments \cite{Aubry_Spectra_I, Tang_Spectra}, a consideration that is of increasing importance to the radar community \cite{Spectral_Coexistence}.  Thus, not only should a MIMO waveform set be adaptive so as to synthesize a desired beampattern, it should also possess these practical properties for operation in real world systems.   

This paper investigates applying the Multi-Tone Sinusoidal Frequency Modulated (MTSFM) waveform model \cite{Hague_AES} to design waveform sets whose correlation matrix closely approximates a desired MIMO transmit beampattern.  The MTSFM waveform possesses an instantaneous phase that is represented as a finite Fourier series.  In previous efforts by the author, these Fourier coefficients were modified to synthesize waveforms with desireable auto and cross correlation properties \cite{Hague_AES, Hague_RadarConf_2020, Hague_Asilomar_2022}.  The adaptive nature of the MTSFM model makes it directly applicable to the MIMO beampattern synthesis problem.   Additionally, the MTSFM waveform model naturally possesses the constant envelope and spectral compactness properties necessary for efficient transmission on practical radar transmitter electronics \cite{Hague_AES}.  These qualities make the MTSFM waveform an intriguing option for MIMO radar systems.  The rest of this paper is organized as follows: Section \ref{sec:sigModel} provides an overview of the MIMO beampattern model as well as the MTSFM waveform model. Section III describes the MIMO beampattern synthesis technique using the MTSFM model and demonstrates it via a well known design example from the literature \cite{MIMO_SanAntonio}.  Finally, Section IV concludes the paper.    

\section{MIMO Waveform Signal Model}
\label{sec:sigModel}
This section describes the MIMO waveform signal model and introduces the MTSFM waveform model.  The MIMO model assumes a Uniform Linear Array (ULA) with $M$ elements and half-wavelength inter-element spacing $d=\lambda/2$.

\subsection{The MIMO Waveform Set}
Each element of the MIMO ULA transmits a unique basebanded waveform $x_m\left(t\right)$ where $m=1,\dots,M$.  Each of the $M$ waveforms in the waveform set are of duration $T$ with time support over the interval $-T/2 \leq t \leq T/2$ expressed as
\begin{equation}
x_m\left(t\right) = \dfrac{\rect\left(t/T\right)}{\sqrt{MT}}e^{j\varphi_m\left(t\right)}
\label{eq:waveformSet}
\end{equation}
where $\varphi_m\left(t\right)$ is the $m^{\text{th}}$ waveform's instantaneous phase and the $1/\sqrt{MT}$ term normalizes each waveform to have a total energy of $1/M$.  The waveform's frequency modulation function, denoted as $f_m\left(t\right)$, maps its instantaneous frequency as a function of time and is expressed as
\begin{equation}
f_m\left(t\right) = \dfrac{1}{2\pi}\dfrac{\partial \varphi_m\left(t\right)}{\partial t}.
\label{eq:modFunc}
\end{equation}
While the primary focus of this paper is the synthesis of waveform sets to realize a desired MIMO beampattern, this paper also briefly discusses the Auto-Ambiguity Function (AAF) characteristics of the synthesized waveforms.  The narrowband AAF measures the correlation between a waveform and its Doppler shifted versions and is expressed as \cite{Levanon, Rihaczek}
\begin{equation}
\chi_{m, m}\left(\tau, \nu\right) =  \int_{-\infty}^{\infty}x_m\left( t-\frac{\tau}{2}\right) x_m^*\left(t+\frac{\tau}{2}\right)e^{j2\pi\nu t} dt
\end{equation}
where $\nu$ is the Doppler shift frequency.  The Auto-Correlation Function (ACF) of the waveform is simply the zero-Doppler cut of the AAF $R\left(\tau\right) = \chi_{m, m}\left(\tau, 0\right)$.  

\subsection{The MIMO Beampattern and Correlation Matrix}
\label{subsec:ulsModel}
The narrowband MIMO transmit beampattern, expressed as a normalized power density at angle $\theta$ and denoted as $P\left(\theta\right)$, is expressed as 
\begin{equation}
P\left(\theta\right) = \mathbf{a}\left(\theta\right)^H \mathbf{R} \mathbf{a}\left(\theta\right)
\label{eq:mimoBeamPattern}
\end{equation}
where $P\left(\theta\right) \geq 0$ and $\mathbf{a}\left(\theta\right)$ is the transmit array steering vector 
\begin{equation}
\mathbf{a}\left(\theta\right) = \left[1,~e^{j\pi \sin\theta},~\dots,~e^{j\pi\left(M-1\right)\sin\theta} \right]^{\text{T}}.
\label{eq:mimoSteerVec}
\end{equation} 
The matrix $\mathbf{R} \in \Re^{M\times M}$ is the MIMO waveform correlation matrix of the waveform set whose entries represent the inner products between all of the waveforms in the set
\begin{equation}
\mathbf{R} = \langle x_m\left(t\right), x_{m'}\left(t\right) \rangle = \int_{-T/2}^{T/2}x_m\left(t\right)x_{m'}^*\left(t\right) dt.
\label{eq:mimoCorrMat}
\end{equation}
For mathematical simplicity, this paper focuses on correlation matrices whose entries are real.  This property produces MIMO beampatterns that are even-symmetric in angle $\theta$ \cite{Dumitrescu}.  

The correlation matrix must possess several characteristics.  The primary condition on \eqref{eq:mimoCorrMat}, which directly follows from \eqref{eq:mimoBeamPattern}, is that it must be positive semi-definite.  Additionally, there is also a constraint that each element transmits equal power.  Using the waveform set model as defined in \eqref{eq:waveformSet} results in the diagonal elements of the correlation matrix \eqref{eq:mimoCorrMat} all being the same value of $1/M$.  The structure of the correlation matrix is completely determined by the relative correlation between each waveform in the waveform set and correspondingly informs the shape of the coherent MIMO transmit beampattern.  If the waveform set is mutually orthogonal, then \eqref{eq:mimoCorrMat} takes the form of a scaled identity matrix $\left(1/M\right)\mathbf{I}_{\text{M}\times\text{M}}$.  This results in an omni-directional transmit beampattern \cite{MIMO_SanAntonio, MIMO_Overview, MIMO_Waveform}.  If all of the waveforms are identical (i.e, a phased array), then all of the entries of \eqref{eq:mimoCorrMat} are $1/M$ and the beampattern resembles that of the standard ULA beampattern shape \cite{MIMO_SanAntonio}.  The design challenge is to synthesize waveform sets that produce MIMO beampattern shapes that lie between these two extremes.  This results in a waveform correlation matrix whose off-diagonal entries are non zero but less than $1/M$ in magnitude.

\subsection{The MTSFM Waveform Model}
\label{subsec:mimoMTSFM}
This paper utilizes the MTSFM waveform model in \cite{Hague_AES} whose modulation function is a finite Fourier series
\begin{equation}
f_m\left(t\right) = \dfrac{a_{m,0}}{2}+\sum_{k=1}^K a_{m,k}\cos\left(\dfrac{2\pi k t}{T}\right)
\label{eq:mtsfmMod}
\end{equation}
where $a_{m,k}$ are the $K$ Fourier coefficients for each of the $M$ waveforms in the set.   For simplicity, the $0^{\text{th}}$ order harmonic coefficients $a_{m,0}$ are all assumed to be 0.  Each waveform in the set $x_m\left(t\right)$ therefore possesses an instantaneous phase $\varphi_m\left(t\right)$ expressed as
\begin{equation}
\varphi_m\left(t\right) = \sum_{k=1}^K \alpha_{m,k} \sin \left(\dfrac{2\pi k t}{T}\right).
\label{eq:MTSFM}
\end{equation}
The $\alpha_{m,k}=a_{m,k}\left(T/k\right)$ terms are the MTSFM waveform set's modulation indices which represent a discrete set of $MK$ design coefficients.  These coefficients are modified to synthesize waveform sets with a correlation matrix $\mathbf{R}$ that realizes a desired MIMO beampattern.  Note that \eqref{eq:mtsfmMod} can also include odd harmonics in its instantaneous phase.  However, this paper focuses solely on waveforms whose modulation functions \eqref{eq:modFunc} are even-symmetric and thus \eqref{eq:mtsfmMod} possesses only cosine harmonics.  

As shown in the appendix, the correlation matrix of a waveform set composed of MTSFM waveforms is expressed in terms of the set of modulation indices $\alpha_{m,k}$ as 
\begin{multline}
\mathbf{R}\left(\{\alpha_{m,k}\}\right)= \left(\dfrac{1}{M}\right) \times \\
  \begin{bmatrix}
    \mathcal{J}_0\left(\{z_{1,1,k}\}\right)		& \mathcal{J}_0\left(\{z_{1,2,k}\}\right)	 & \dots   & \mathcal{J}_0\left(\{z_{1,M,k}\}\right)	 \\
    \mathcal{J}_0\left(\{z_{2,1,k}\}\right)	 & \mathcal{J}_0\left(\{z_{2,2,k}\}\right)	& \dots   & 		    \mathcal{J}_0\left(\{z_{2,M,k}\}\right)	 \\
    \vdots 		               &                              & \ddots & \vdots \\
    \mathcal{J}_0\left(\{z_{M,1,k}\}\right) & \mathcal{J}_0\left(\{z_{M,2,k}\}\right)		 & \dots   & \mathcal{J}_0\left(\{z_{M,M,k}\}\right)	
  \end{bmatrix}
  \label{eq:rMat}
\end{multline}
where $\mathcal{J}_0\left(\{z_{m,m',k}\}\right)$ is the $0^{\text{th}}$ order cylindrical Generalized Bessel Function (GBF) \cite{Dattoli} with $K$-dimensional argument $z_{m,m',k} = \alpha_{m,k}-\alpha_{m',k}$.  The cylindrical GBF is a purely real-valued function of the $K$-dimensional arguments resulting in a real-symmetric correlation matrix.  This in turn will produce and even-symmetric MIMO beampattern $P\left(\theta\right)$.  

\section{Synthesis of MTSFM MIMO Waveform Sets}
\label{sec:mtsfmMIMO}
This section describes the methods used to optimize MTSFM MIMO waveform sets to approximate a desired beampattern shape.  It then demonstrates this waveform design method using a well known design example from the literature.  

\subsection{Design Methodology}
\label{subsec:methods}
This work utilizes a modified version of the cost function defined in \cite{MIMO_SanAntonio} to synthesize MTSFM waveform sets \eqref{eq:MTSFM} to approximate a desired MIMO transmit beampattern.  The optimization problem is formally expressed as
\begin{multline}
\underset{\alpha_{m,k}}{\text{min}}\int_{\theta} \Bigl | P_d\left(\theta\right)-\mathbf{a}\left(\theta\right)^H\mathbf{R}\left(\{\alpha_{m,k}\}\right)\mathbf{a}\left(\theta\right) \Bigr |^2\cos\theta d\theta \\ \text{s.t.~} \beta_{rms}^2\left(\{\alpha_{m,k}\}\right) \in \left(1\pm\delta\right)\beta_{rms}^2(\{\alpha_{m,k}^{\left(0\right)}\})
\label{eq:costFunction}
\end{multline}
where $P_d\left(\theta\right)$ is the desired beampattern shape and $\mathbf{R}\left(\{\alpha_{m,k}\}\right)$ is the correlation matrix of the MTSFM waveform set \eqref{eq:rMat}.  The quantity $\beta_{rms}^2\left(\{\alpha_{m,k}\}\right)$ is the Root-Mean Square (RMS) bandwidth of each waveform in the set.  The RMS bandwidth measures the spread of the waveform's spectra about its spectral centroid \cite{Cohen} and for the MTSFM waveform this is expressed as \cite{Hague_AES}
\begin{equation}
\beta_{rms}^2\left(\{\alpha_{m,k}\}\right) = \left(\frac{2\pi}{T}\right)^2\sum_{k=1}^K k^2 \dfrac{\alpha_{m,k}^2}{2}.
\end{equation}  The intent of the RMS bandwidth constraint is to ensure that the RMS bandwidths of the optimized waveforms in the set are within some tolerance $\delta$ of the RMS bandwidths of the initial waveforms $\beta_{rms}^2(\{\alpha_{m,k}^{\left(0\right)}\})$.  This constraint acts to preserve the spectral compactness inherent in the initial waveform set that is passed to the optimization routine.  The optimization problem \eqref{eq:costFunction} is numerically evaluated using Matlab's optimization toolbox function \emph{fmincon} \cite{Matlab}.  

\begin{figure}[ht]
\centering
\includegraphics[width=0.5\textwidth]{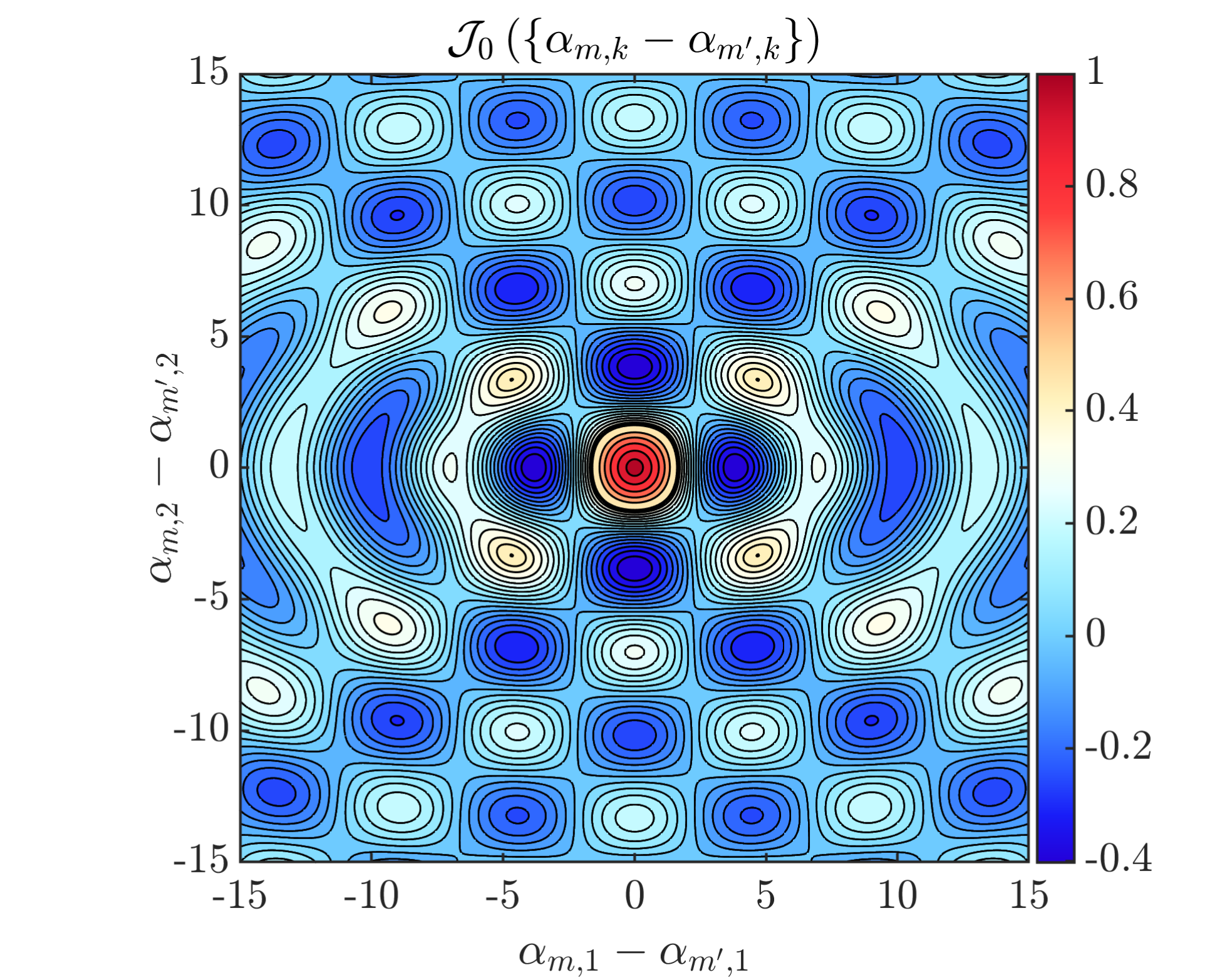}
\caption{Plot of a $0^{\text{th}}$ order cylindrical GBF with two-dimensional argument $\alpha_{m,k}-\alpha_{m',k}$ representing a single entry of the MTSFM waveform set's correlation matrix.  The GBFs are highly oscillatory functions with multiple local extrema.  Optimizing a MTSFM waveform set's MIMO beampattern results in a multi-modal objective function with no clear guarantees on finding a globally optimum value.}
\label{fig:GBF}
\end{figure}

The MIMO beampattern realized by a MTSFM waveform set is a function of its correlation matrix whose elements are given by the $0^{\text{th}}$ order cylindrical GBF as shown in \eqref{eq:rMat}. Figure \ref{fig:GBF} shows a 2-D cylindrical GBF for illustrative purposes.  Much like their 1-D counterparts, the GBFs are highly oscillatory functions across their $K$-dimensional arguments with multiple local extrema.  Thus, the objective function defined in \eqref{eq:costFunction} will also exhibit oscillatory behavior across the parameter space $\alpha_{m,k}$.  This results in a highly multi-modal objective surface where finding a global minimum is unlikely.  Thus, the optimization problem \eqref{eq:costFunction} is run over 100 trials each with a different initial set of MTSFM modulation indices to fully explore the space of possible approximations to the desired MIMO beampattern $P_d\left(\theta\right)$ using MTSFM waveform sets.  For each trial, each of the $M$ sets of $K$ modulation indices $\alpha_{m,k}$ are all set equal to each other and then perturbed with i.i.d Gaussian noise with variance $\sigma^2=1/100$.  This produces a waveform set with almost perfect correlation which closely approaches a phased-array configuration.  Experimental evaluations of \eqref{eq:costFunction} showed it was better to initialize with strongly correlated waveforms and reduce their mutual correlation rather than to initialize with uncorrelated waveforms and attempt to increase their mutual correlation.    

\subsection{An Illustrative Design Example}
\label{subsec:example}

\begin{figure}[ht]
\centering
\includegraphics[width=0.5\textwidth]{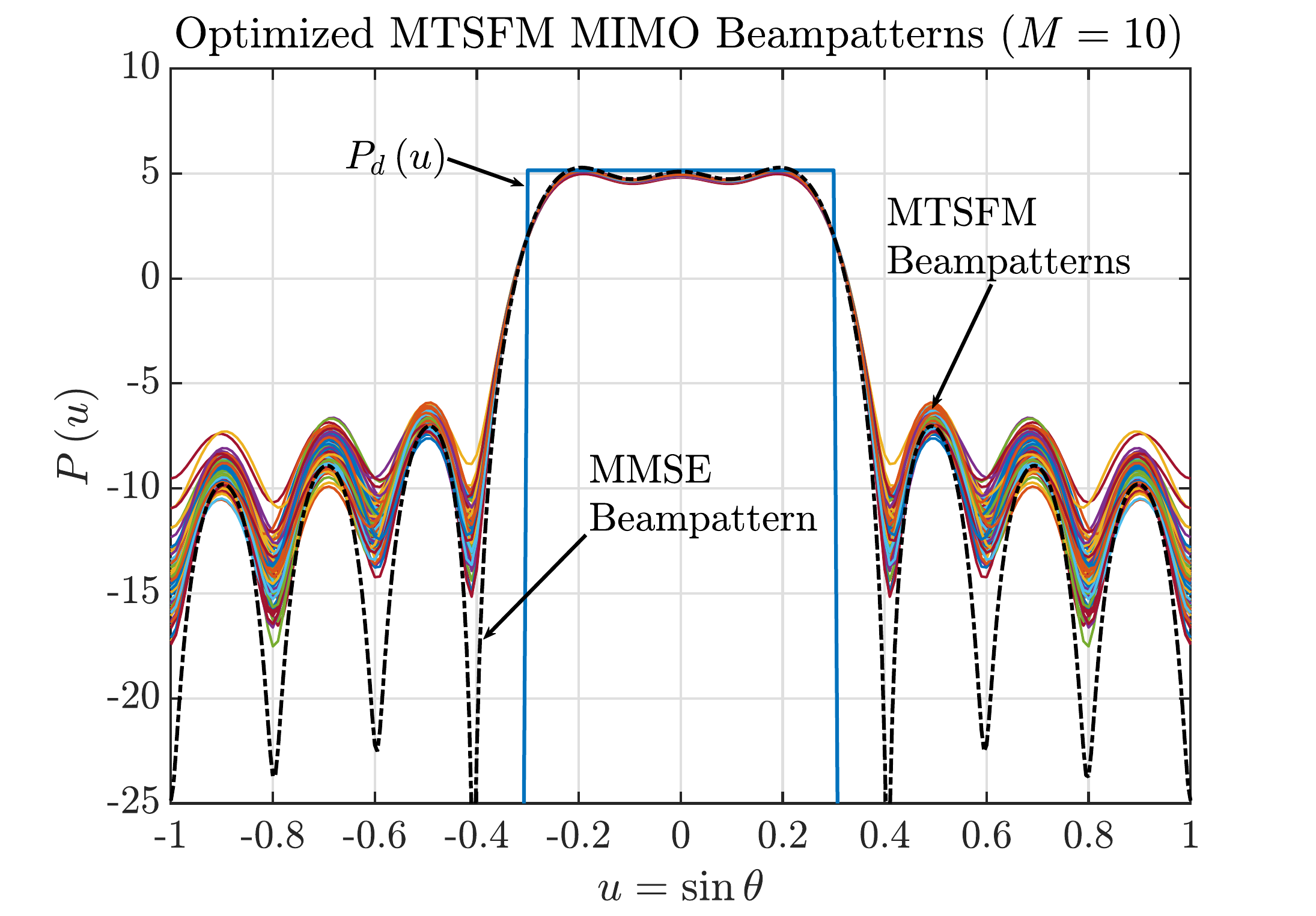}
\caption{Desired MIMO Beampattern $P_d\left(u\right)$ and synthesized MTSFM MIMO beampatterns utilizing 100 distinct MTSFM waveform sets each of size $M=10$.  Also shown in the black dashed line is the MMSE MIMO beampattern using the algorithm developed in \cite{MIMO_SanAntonio}.  Each MTSFM waveform set was optimized using \eqref{eq:costFunction} with $\delta=0.1$.  The adaptability of the MTSFM waveform model facilitates designing a large number of distinct waveform sets that closely approximate a desired MIMO transmit beampattern.}
\label{fig:MIMO_1}
\end{figure}

The following design example demonstrates the ability of the MTSFM waveform model to synthesize unique MIMO beampatterns. This design example is based off of the one shown in Figure 6 of \cite{MIMO_SanAntonio} which uses an ULA with $M=10$ elements and seeks to approximate the desired beampattern $P_d\left(\theta\right)$ 
\begin{equation}  P_d\left(\theta\right) = \left\{
\begin{array}{ll}
	A, & |u=\sin\theta| \leq 0.3\\

      0, & owise\\
\end{array} 
\right.
\label{eq:region} 
\end{equation}
using \eqref{eq:costFunction} where $A$ is a normalization constant that ensures the total power across $\theta$ equals the dimensionality of the array $M$ \cite{MIMO_SanAntonio}.   Figure \ref{fig:MIMO_1} shows the MIMO beampatterns as a function of $u=\sin\theta$ of 100 distinct optimized MTSFM waveform sets each composed of $M=10$ waveforms.  The MTSFM waveform sets were optimized using \eqref{eq:costFunction} with $\delta = 0.1$.  Also shown in the figure is $P_d\left(u\right)$ and the MMSE beampattern achieved by solving \eqref{eq:costFunction} using the algorithm described in \cite{MIMO_SanAntonio}.  Each waveform in the set possesses $K=16$ modulation indices and a time-bandwidth product $T\Delta f = 64$.  From the figure, it is clear the resulting MIMO beampatterns bear strong resemblance to the MMSE beampattern placing the majority of their power in the passband region of \eqref{eq:region} with very little variation across $u$.  Additionally, these beampatterns possess reasonably low sidelobes outside that region with the median Peak-to-Sidelobe Level Ratio (PSLR) being -11.89 dB compared to the -12.31 dB PSLR achieved using the MMSE algorithm from \cite{MIMO_SanAntonio}.  The lowest PSLR achieved by any MTSFM waveform set was -12.63 dB and the highest achieved was -11.00 dB.

\subsection{Spectral and AAF Characteristics of the MTSFM Waveform Sets}
\label{subsec:AAF}
\begin{figure}[ht]
\centering
\includegraphics[width=0.5\textwidth]{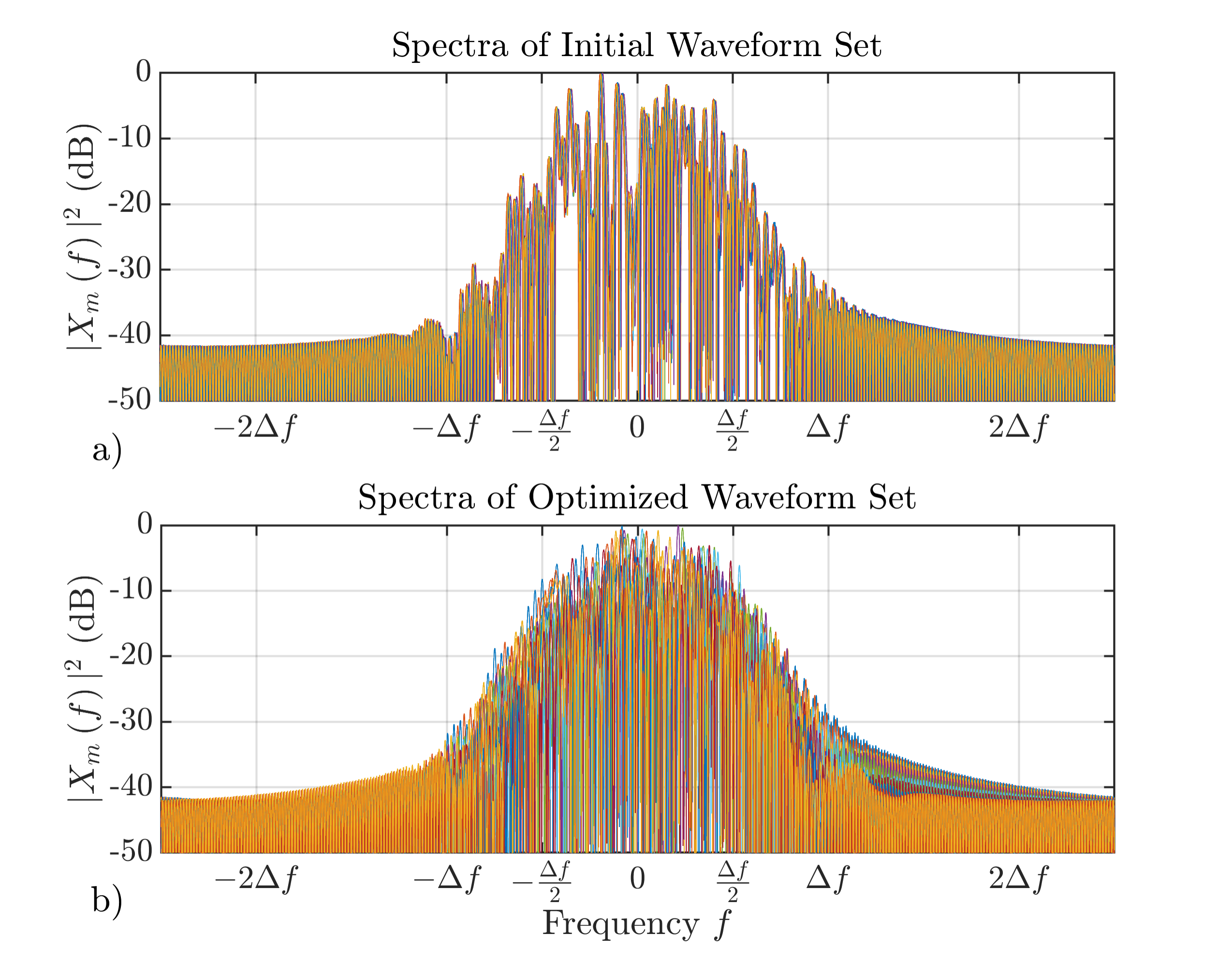}
\caption{Magnitude squared spectra $|X_m\left(f\right)|^2$ of an initial waveform set (a) and its corresponding optimized (b) set resulting from running the optimization routine \eqref{eq:costFunction}.  All of the waveforms from both waveform sets possess a compact spectral shape.}
\label{fig:mtsfmSpectra}
\end{figure}

The results in Figure \ref{fig:MIMO_1} demonstrate the MTSFM waveform's ability to consistently realize a desired beampattern with diverse waveform sets.  It is also insightful to examine the spectral and AAF properties of the synthesized MTSFM waveform sets.  Figure \ref{fig:mtsfmSpectra} shows the magnitude squared spectra $|X_m\left(f\right)|^2$ of all the waveforms from one of the initial and optimized waveform sets discussed in the previous section.  Both sets of waveforms possess spectra with the vast majority of their energy being densely concentrated in their swept bandwidths $\Delta f$.  Coupling this with the natural constant modulus property of FM waveforms makes the MTSFM MIMO waveform sets well suited for efficient transmission on practical radar transmitter electronics.

\begin{figure}[ht]
\centering
\includegraphics[width=0.5\textwidth]{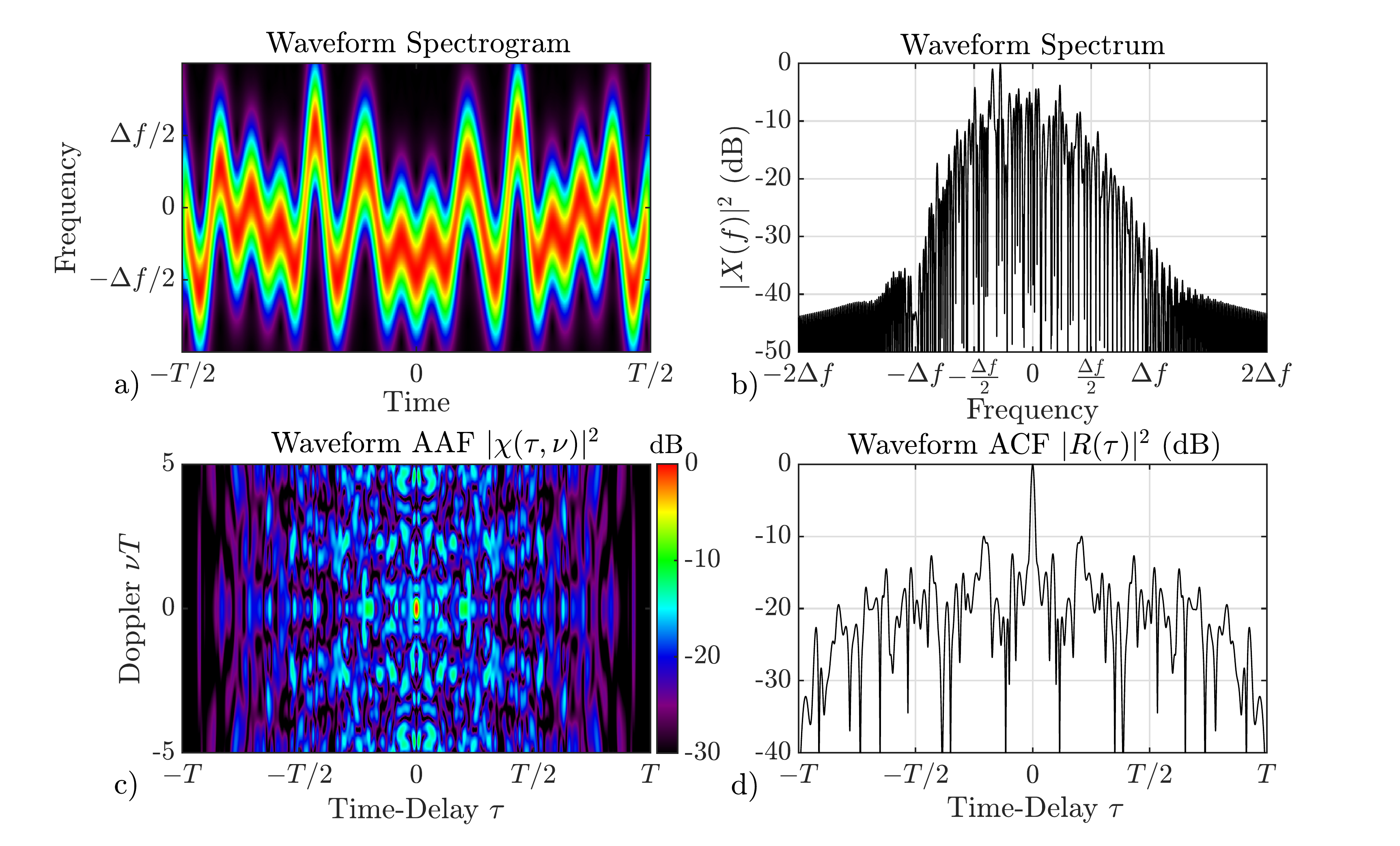}
\caption{Spectrogram (a), spectrum (b), AAF (c), and ACF (d) of one of the MTSFM waveforms from the optimized MIMO waveform set shown in Figure 3.  All MTSFM waveforms from both the initial and optimized waveform sets possess a ``Thumbtack-Like'' AAF shape.}
\label{fig:mtsfmAAF}
\end{figure}

Figure \ref{fig:mtsfmAAF} shows the spectrogram, spectrum, AAF, and ACF of one of the MTSFM waveforms from the optimized waveform set.  As can be seen in the spectrogram, the MTSFM's frequency modulation function changes smoothly with time which results in its compact spectral shape \cite{Hague_AES}.  This particular variant of the MTSFM possesses an even-symmetric modulation function as defined in \eqref{eq:mtsfmMod}.  This results in the waveform having a ``Thumbtack-Like'' AAF shape \cite{Rihaczek} with an uncoupled mainlobe \cite{Hague_UACE}.  The rest of the bounded AAF volume is spread evenly in $\tau$ and $\nu$ resulting in a pedestal of sidelobes whose height is inversely proportional to the waveform's time-bandwidth product.  The sidelobe pedestal can be clearly seen in the ACF plot in Figure \ref{fig:mtsfmAAF}.  All other waveforms in both the initial and optimized waveform sets exhibited strongly similar AAF/ACF characteristics.  

\section{Conclusion}
\label{sec:Conclusion}
This paper investigated applying the MTSFM waveform model to the MIMO beampattern synthesis problem.  The MTSFM waveform model allows for synthesizing waveform sets that closely approximate a desired MIMO beampattern while also possessing the constant modulus and spectral compactness properties necessary for efficient transmission on practical radar transmit electronics.  Each waveform in the waveform set attains a ``Thumbtack-Like'' AAF shape.  Future efforts will focus on extending analysis to MTSFM waveforms with odd symmetry in their modulation functions, extending the model to the broadband regime in a manner similar to \cite{MIMO_SanAntonio_Broadband}, and designing MTSFM waveform sets with ripple control in the beampattern shape.

\appendices

\section{MTSFM MIMO Correlation Matrix}
\label{sec:AppendixI}
The inner product between any two MTSFM waveforms in a set of $M$ waveforms is expressed as 
\begin{equation}
\langle x_m\left(t\right), x_{m'}\left(t\right) \rangle = \int_{-T/2}^{T/2}x_m\left(t\right) x^*_{m'}\left(t\right) dt.
\label{eq:innerProd}
\end{equation}
Inserting \eqref{eq:MTSFM} into \eqref{eq:waveformSet}, the integrand in \eqref{eq:innerProd} is then expressed as 
\begin{equation}
\dfrac{1}{MT}\exp\left[j\sum_{k=1}^K \left(\alpha_{m,k}-\alpha_{m',k}\right)\sin\left(\dfrac{2\pi k t}{T}\right) \right].
\label{eq:integrand1}
\end{equation}
Utilizing the Jacobi-Anger expansion for GBFs \cite{Dattoli}, \eqref{eq:integrand1} simplifies to
\begin{equation}
\dfrac{1}{MT} \sum_{n=-\infty}^{\infty} \mathcal{J}_n^{1:K}\left(\{\alpha_{m,k}-\alpha_{m',k}\}\right) e^{j\frac{2\pi n t}{T}}
\label{eq:integrand2}
\end{equation}
where $\mathcal{J}_n^{1:K}\left(\{\alpha_{m,k}-\alpha_{m',k}\}\right)$ is the $n^{\text{th}}$ order $K$-dimensional cylindrical GBF with argument $\{\alpha_{m,k}-\alpha_{m',k}\}$.  Inserting \eqref{eq:integrand2} into \eqref{eq:innerProd} results in the integral expression
\begin{equation}
\dfrac{1}{MT} \sum_{n=-\infty}^{\infty} \mathcal{J}_n^{1:K}\left(\{\alpha_{m,k}-\alpha_{m',k}\}\right) \int_{-T/2}^{T/2}e^{j\frac{2\pi n t}{T}}dt.
\label{eq:integrand3}
\end{equation}
The integral in \eqref{eq:integrand3} evaluates to $T\sinc\left[\pi n\right]$ which is nonzero only when $n=0$.  Thus, the inner product between any two MTSFM waveforms is expressed as
\begin{equation}
\langle x_m\left(t\right), x_{m'}\left(t\right) \rangle = \dfrac{1}{M} \mathcal{J}_0^{1:K}\left(\{\alpha_{m,k}-\alpha_{m',k}\}\right) 
\end{equation}


\end{document}